# Topological photonics with scattering media

E. Cusworth[1], V. G. Kravets[1], F. Fraggelakis[2], P. Lingos[2], E. Stratakis[2], A. V. Kabashin[3], A. N. Grigorenko[1*]

[1] *Department of Physics and Astronomy, University of Manchester, Manchester, M13 9PL, UK*

[2] *Department of Physics, University of Crete, 700 13 Heraklion, Crete, Greece*

[3] *Aix Marseille University, CNRS, UMR 7341 CNRS, LP3, Campus de Luminy – case 917, 13288, Marseille Cedex 9, France*

**Abstract:** Scattering media, being ubiquitous in nature and critically important for assessments (e.g., biological tissues), are often considered as nuisance in optics. Here we show that it is not always the case and scattering media could be essential in providing elements of topological photonics. We demonstrate that topological darkness can be realised in the presence of scattering media due to the transverse nature of electromagnetic waves and the hairy ball theorem. We describe realizations of topological darkness in a scattering medium comprising composite gold nanostructures produced by a low-cost technique of laser structuring of thin metal films. Our results can be useful for a variety of tasks, including topological nanophotonics, optical label-free biosensing, and biomedical photonics.

* Corresponding author. Email: sasha@manchester.ac.uk

Scattering media are omnipresent in nature: biological tissues, planetary and stellar atmospheres, milk and oceans are all examples of scattering media. Study of light propagation in scattering media had started in works of Rayleigh (*1*), Mie (*2*), Chandrasekhar (*3*) and continued by Rytov (*4*), Lagendijk (*5*) and others (*6*). For a long time, scattering media were considered to be a hindrance to optics and several ingenious methods were developed to overcome scattering. Fujimoto *et. al.* suggested optical coherence tomography that allowed one to perform non-invasive imaging in biological systems (*7, 8*) and introduced optical coherence microscopy that permitted imaging deep into highly scattering media (*9, 10*). These techniques were modified to image fluorescent objects hidden behind opaque scattering layers (*11*).

It was found later that, instead of being considered as hindrance, a scattering medium could be an important element of optical devices that provide better imaging and focusing of light with the help of suitable reconstruction of light wave-fronts (*12-15*), for review see (*16*). A wave-front shaping technique, designed to correct effects of scattering, was also applied for focusing and compression of ultrashort light pulses (*17*).

Here we extend the idea that scattering media could be useful in optical methods and devices by utilising topology of light phase (*18*) and polarization (*19*). Namely, we prove a proposition that for an optical device with a scattering medium illuminated at fixed angle of light incidence and fixed incident light wavelength there exists at least one direction with topologically protected zero scattering (total darkness) and corresponding singular (Heaviside-like) behaviour of light phase (*20*) due to simple topological arguments connected to the transverse nature of electromagnetic waves. The directions of zero scattering and related phase singularities are important for the implementation of components of topological photonics. They could also be used in critically important applications, such as ultrasensitive low-cost optical biosensing (*20, 21*) and high-resolution optical diagnose of biological tissues and histological samples (*22, 23*) benefiting considerably from extreme phase changes near the points of darkness. We experimentally demonstrate the concept of topological darkness (TD) in scattering using structures made of laser-structured thin gold films. Our work will enable the use of various scattering media in optical devices based on light topology. It can be considered as a natural extension of the concept of topologically protected light nodes observed in real space(*24, 25*) to the Fourier *k*-space, for a recent review see (*26*).

TD is a phenomenon where reflection (or transmission) of light from an optical hetero/nanostructure can be made exactly zero due to topology of intersection of the zero reflection (or transmission) surface derived from Fresnel coefficients with the spectral parameter curve of an interrogated layer insured by the Jordan-Brouwer theorem (*20, 27-30*). Topological darkness generates abrupt phase changes due to singular behaviour of light phase (the Heaviside jumps of the phase by 180°) at the points of darkness (*14, 18, 20, 21, 30*). One of the most important features of TD is its robustness against small imperfections in a metamaterial designed to achieve TD. These small imperfections do not destroy zero reflection/transmission (total darkness); instead, zero reflection/transmission will be observed at slightly different angle of incidence and wavelength of incident light (*20, 30*). However, when imperfections (irregularities in a structure) in the studied medium are large, the material behaves as a scattering medium that cannot be described by effective optical constants and Fresnel theory; therefore, the topological considerations presented in Ref. (*20*) cannot be applied. Indeed, a scattering medium normally produces light rays in all possible directions (these directions are not restricted to those of reflection and refraction).

It is worth stressing that a scattering medium is a standard medium that is used in biological applications. Biological tissues, blood, DNA assemblies are all scattering media. It requires a lot of time and effort to arrange biological objects in crystalline forms as the ingenious efforts of Rosalind Franklin showed (*31*). When the number of scattering directions is limited, the optical properties of an interrogated medium can be described by a Fourier series. We recently showed that for such materials – which we termed as Fourier metamaterials (*32*) – TD can be observed not only in reflection but also in other Fourier directions. Phase singularities associated with darkness were also discussed by the names of wave dislocations (*18*), topological charges (*26*), point singularities (*33, 34*); while Fourier nanotransducers based on such metamaterials have recently attracted attention under a different guise of diffractive non-local metasurfaces (*35*) or metagratings (*36*).

In order to demonstrate that TD is still possible for a general case of a scattering object with large irregularities (where light scattering is described by a Fourier integral instead of a Fourier series), we prove a simple proposition: any reasonably small-sized scattering object illuminated by monochromatic light possesses scattering directions protected by topology in which the scattering light amplitude is exactly zero provided all scattering fields are linearly polarized. Hence, TD can also be observed for such scattering media at some angles of scattering. To prove this proposition, let us consider a small object that scatters light in all possible directions. Figure 1 shows this object as a yellow disk, the incident light (green arrow) is denoted by 0-index, blue arrows show two possible scattered electromagnetic waves marked by prime indices.

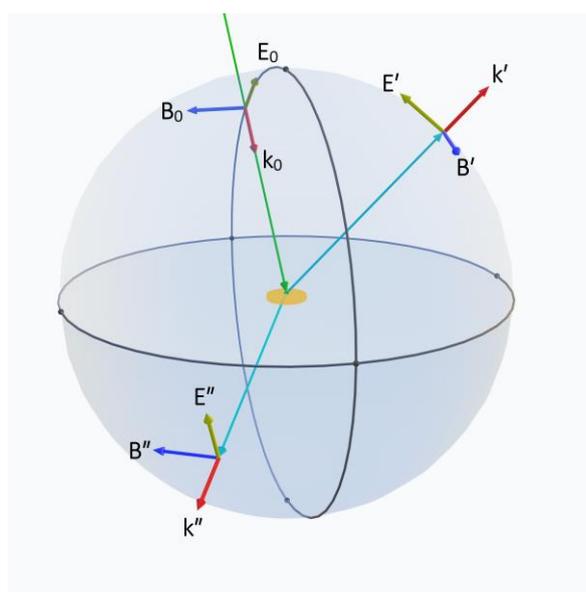

Figure 1. Topological darkness for a scattering medium.

Since the scattering object is small, we can assume that the polarization of the scattered light in any given direction is transverse and hence an electric field of the scattered electromagnetic wave is tangential to the surface of the sphere. Then, the existence of TD in scattering follows directly from the hairy ball theorem which states that there is no non-vanishing continuous tangent vector field on the ordinary 2D sphere (*37*). As a result, any reasonably small scattering object (such that its scattering fields are linearly polarized) illuminated by light has a direction in which scattered light amplitude is exactly zero due to topological arguments. This allows us to extend the concept of TD to scattering objects, materials and media which are not described by Fresnel laws of light reflection and refraction. It is worth mentioning that analogous considerations were previously applied to Mie scattering (*38*) and electromagnetic multipole scattering (*39*) with the same conclusion of the absence of scattering for some isolated directions based on the hairy ball theorem in the case of linearly polarized scattering fields. There are many objects/media with linearly polarized scattering. Most of optical media that does not contain chiral structures realise exactly this type of scattering being illuminated by linear polarized light. A prominent example of such scenario is Rayleigh scattering. (In parenthesis we note that the case of elliptically polarized scattering fields is more complicated. At each given moment of time we can apply arguments of Figure 1 even in this case. However, the electric field vectors are rotating with time for elliptical polarization and hence the direction of TD could change with time (*38*). In this case there exists another topological invariant: the directions in which the scattered light is exactly circularly polarized due to Hopf invariants, see discussion in (*38, 39*). This case is realised when a medium is illuminated by elliptically polarized light or has chiral structures inside.)

To experimentally demonstrate TD in scattering media, we have employed a new type of metasurface (a disordered low-amplitude grating) produced by laser structuring of thin metal films deposited on a dielectric substrate (*40*). It is well known that laser processing of thick layers of materials can lead to the creation of laser induced periodic surface structures (LIPSS) (*41*). Laser structuring of thin metal films is much less studied but it could also lead to LIPSS of different periodicities. Recently, we found that femtosecond laser structuring of thin gold films (depending on the fabrication parameters) could lead to either well-defined periodic structures or to structures with reasonably high disorder due to the presence of several LIPSS (*40*).

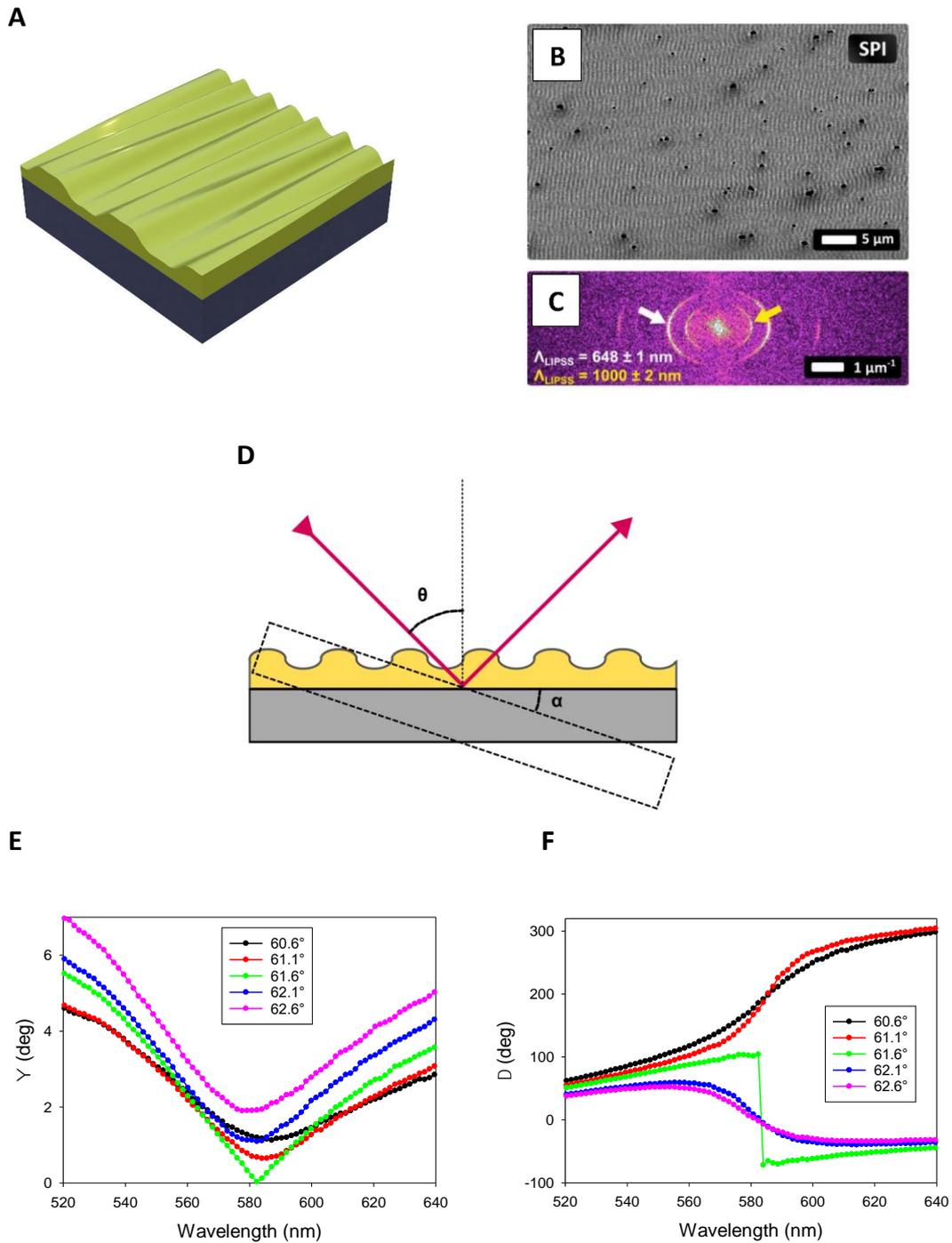

Figure 2. Topological darkness for LIPSS. (**A**) 3D schematics of laser structured thin gold films (yellow colour) on a dielectric substrate (blue colour). (**B**) SEM images of non-periodic profiles produced by laser structuring. (**C**) Fourier transform of the map shown in (B) that confirms the presence of scattering. (**D**) Geometry of scattering measurements. $\theta$ is the angle of incidence, $\alpha$ is the tilt of the substrate. (**E**) TD observed at angle of incidence 61.6° for standard reflection (where $\alpha$ = 0). $\Psi$ represents the ellipsometric reflection, see Supplementary Materials. (**F**) An angular dependence of ellipsometric phase $\Delta$ spectra at TD (notice the Heaviside-like jump of phase by 180° at the point of TD for angle of incidence of 61.6°).

Figure 2A depicts a schematic of the morphology produced by femtosecond laser structuring of a thin (32nm) gold film on a glass substrate, while Fig. 2B shows a typical scanning electron microscopy image of a laser-processed sample. Due to a competition between two different LIPSS periods arising in thin film geometry (see Fig. 2C and Supplementary Materials (SM)), a fabricated structure has a large level of disorder and scatters light not only in the directions of light reflection but also in other directions. As a result, we were able to detect significant light intensity at angles close to the direction of reflection but not equal to it. Figure 2D shows the measurement geometry – the light reflection from our samples was measured with the help of a variable angle spectroscopic ellipsometer (J. A. Woollam M2000F) – at the geometry of input and output legs corresponding to the standard reflection measurements at some angle $\theta$ in the case where the sample plane was tilted (in-plane) by some angle $\alpha$ (Fig. 2D). It is worth noting that there are two tilting angles for a sample (the in-plane and the out-of-plane tilts) and, for the sake of demonstration, we present data only for changes of the in-plane tilt angle $\alpha$. The second tilt angle and the orientation of the sample in the plane were chosen in such a way that the crosstalk between *p*- and *s*-polarizations was absent, so that $S_{ps} \approx 0$, where $S_{ps}$ is the scattering in *s*-polarization measured under illumination of the sample by *p*-polarized light. This implies that, at our conditions, the sample illuminated by only *p*-polarization produced only *p*-polarized scattered fields. For a tilted sample, angle of light incidence is equal to $\theta + \alpha$ while the scattering is measured at angle of $\theta - \alpha$. (At the zero-tilt angle, $\alpha = 0$, we were measuring the standard reflection from a sample at angle of incidence $\theta$).

The first surprising result is shown in Fig. 2E, F. Despite a large amount of scattering, laser-structured samples showed TD in the standard reflection geometry where the tilt angle $\alpha$ was zero. Figure 2E demonstrates the evolution of the spectrum of the ellipsometric reflection measured from a fabricated sample under different angles of incidence (the definition of ellipsometric reflection and phase and its connection to intensity reflection is given in SM). One can see that reflection from the sample goes to extremely low values at the angle of incidence of 61.6° and the wavelength of around 583 nm. Figure 2F confirms that this extremely low reflection is indeed due to TD by plotting the evolution of the phase spectra with the Heaviside-like $\pi$-jump of the light phase observed for $\theta$=61.6° and $\lambda$=583 nm (see also a 3D plot of the light phase $\Delta$ in SM). Such behaviour of the light phase comes from topological properties of darkness as explained in (*14, 20*). Hence, when the sample was illuminated by light of *p*-polarization both components of scattering electric field were zero at some conditions ($S_p$=0 at $\lambda$=583 nm since $\Psi$=0 and $S_{ps}$=0 due to the choice of the sample orientation and the second tilt angle) which indicates complete darkness for our scattering samples for given direction and wavelength. As we mentioned above, the existence of TD for scattering media is not guaranteed by the Jordan-Brouwer theorem (*20, 28, 29*) due to breakdown of Fresnel formulae for such media. More details on scattering of our samples near TD with spectra of scattering for *p*- and *s*-polarizations are given in Supplementary Materials.

The second striking result is the fact that we can observe TD for any angle of incidence near TD in reflection by adjusting the tilt angle $\alpha$. This is in contrast with TD for non-scattering media described by the Fresnel theory where the angle and the wavelength of TD are fixed by intersection of the material constant curve with the zero reflection surface (*42, 43*). For a scattering medium, however, general considerations presented in Fig. 1 suggest that TD can be observed for any angle of incidence by detecting light at some scattering direction. Figure 3A shows that this is indeed the case by demonstrating TD in scattering for two angles near the angle of TD in reflection of $\theta$=61.6°. To achieve total darkness the tilt angle $\alpha$ was adjusted so that angle of incidence for the light was different from angle of detection. For example, the angle of incidence for the green curve of Fig. 3A was $\theta + \alpha$ = 61.1° and TD was measured at the scattering angle of $\theta - \alpha$ = 63.1°, while the angle of incidence for the black curve of Fig. 3A was 62.3° and TD was measured at the scattering angle of 59.5°. (It is worth noting that TD was observed for a whole set of angles around $\theta$=61.6°. We have chosen to show just two angles for clarity. Crosstalk between *p*- and *s*-polarizations was negligible for all these measurements.) The Heaviside-like $\pi$-jumps of the light phase (corresponding to TD of Fig. 3A) are shown in Fig. 3B and confirm the presence of TD. To add another proof that we indeed observe the singular phase behaviour in scattering, we have plotted the dependence of light phase for $\theta$=62.1° as a function of light wavelength and the sample tilt $\alpha$ in Fig. 3C. We can clearly see that there is only one tilt angle that generates the Heaviside-like jump of the phase by exactly $\pi$ radians.

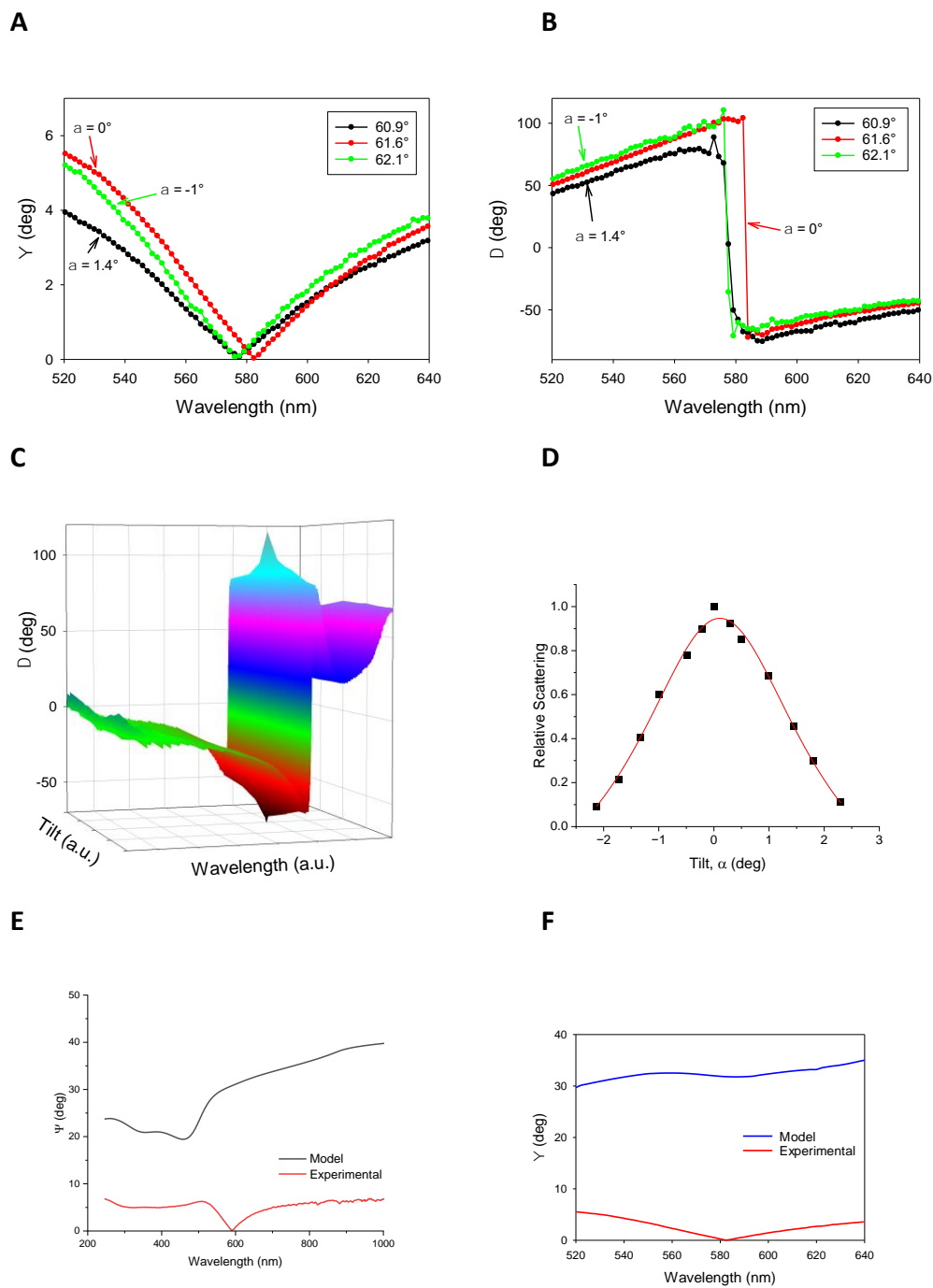

Figure 3. Topological darkness in scattering. (**A**) TD observed in scattering for the angle $\theta$ = 60.9° and the tilt angle $\alpha$ = 1.4°, the black curve, $\theta$ = 61.6° $\alpha$ = 0°, the red curve, $\theta$ = 62.1° $\alpha$ = -1°, the green curve. (**B**) The Heaviside-like jumps of the phase by 180° for TD shown in (A). (**C**) The evolution of the phase spectra with the tilt angle $\alpha$ for $\theta$ = 62.1°. (**D**) Intensity of the scattered light at $\theta$ = 61.6° as a function of the tilt angle $\alpha$. The red curve shows the fitting. (**E**) Comparison of the experimental ellipsometric reflection $\Psi$, the red curve, with the modelling based on the Fresnel theory for the studied sample, the black curve. The inset shows the geometry used in the model. (**F**) Modelling with the help of rigorous coupled wave analysis for the data measured at $\theta$ = 61.6° and the tilt $\alpha$ = 0°.

To be useful in practical applications (e.g., optical biomedical imaging or label-free biosensing), the light intensity near the points of TD should be of sufficient intensity. We found that this is indeed the case. Figure 3D plots the dependence of light scattering intensity as a function of the tilt angle $\alpha$ for fixed angle $\theta$=61.6°. Taking into account the fact that our detection device (a J. A. Woollam ellipsometer) was able to measure the light intensity at the level of 0.5% of incident light intensity, we conclude the changes of $\alpha$ between -2° to 2° were easily detectable (the scattering intensity dropped to 10% as compared to the zero tilt angle, see Fig. 3D) and can be used for practical applications (e.g., ultrasensitive phase biosensing). It is worth also noting that by using the out-of-plane tilt angle we can achieve TD in scattering for any given wavelength and any given incident angle (near the parameters of TD in reflection) as our general proposition (Fig. 1) proves.

Finally, we briefly discuss theoretical modelling of TD in scattering. It is clear that an application of the Fresnel theory to our samples cannot explain the measured data due to the presence of scattering. Figure 3E shows typical Fresnel-based calculations with the effective medium layer replacing the nanostructured gold on a thin non-ablated gold layer (for details see (*40*) and SM) which fails to reproduce the main features of TD observed in our samples. Scattering could be taken into account by rigorous coupled wave approximation (RCWA), Fig. 3F. Unfortunately, RCWA works well only for periodic structures. Since our structures are not exactly periodic, the modelled response does not describe the measured data either, see Fig. 3F and SM. This implies that modelling of TD in scattering requires new theoretical methodologies.

In conclusion, we have demonstrated that TD can be observed for a scattering medium. The topological arguments for this darkness are based on the hairy ball theorem and are completely different from the topological arguments for TD of non-scattering structures described by the Fresnel theory which are based on the Jordan-Brouwer theorem. We have employed a new type of metasurface (a disordered low-amplitude grating) produced by femtosecond laser ablation of thin metal films to experimentally realise TD in scattering. Our results suggest that any medium with strong scattering can be used for elements of topological photonics and could offer an appealing pathway toward the advancement of diverse applications. As one of such applications, we envision ultra-sensitive label-free biosensing (*44*). The implementation of most performing phase-sensitive plasmonic biosensing transducers typically requires highly-ordered metamaterial arrays, which are fabricated by very expensive and time-consuming electron beam lithography or focused ion beam methods (*45*). As we showed, topological darkness and related phase singularities can be observed with essentially disordered nanostructures produced by very cheap and scalable technique of laser processing of thin films, which promises a radical reduction of the cost of ultrasensitive optical biosensing technology. As another application, we see biomedical photonics, as the proposed approach can be applied to enhance the resolution of images in optical characterization of highly scattering biological tissues and histological samples.

**Acknowledgements**: A.N.G and V.G.K. acknowledge the support of Graphene Flagship programme, Core 3 (881603). E. C. acknowledges support of NOWNANO CDT programme funded by EPSRC grant EP/L01548X/1. A.V.K. acknowledges support from the French government under the France 2030 investment plan, as part of the Initiative d'Excellence d'Aix-Marseille Université – A*MIDEX " AMX-22-RE-AB-107. F.F., E.S. and A.V.K. acknowledge contribution of the International Associated Laboratory (LIA) "Laser Matter Interaction from fundamental studies to inNOvative laSer processing- MINOS".

**Author contributions**: A.N.G. conceived the idea of the project. F.F., A.V.K. and E.S. performed sample fabrications and characterizations. V.G.K. performed optical measurements. E.C. performed theoretical modelling and analysed the data. A.N.G. supervised the project. All authors contributed to the writing of the manuscript.

**Competing Interest**: The authors declare no competing interests.

**Data and materials availability**: Experimental and theoretical data are available under a reasonable request.

**SUPPLEMENTARY MATERIALS**: can be found on the website. They provide details of sample fabrications and theory involved.

**References.**